\documentclass[prb,showpacs,aps,twocolumn]{revtex4}
\usepackage{amsmath}
\usepackage{graphicx}
\usepackage{dcolumn}
\usepackage{amssymb}

\begin{document}

\title{Cyclotron resonance of a magnetic quantum dot}
\author{Nga T. T. Nguyen}
\email{nga.nguyen@ua.ac.be}
\author{F. M. Peeters}
\email{francois.peeters@ua.ac.be} \affiliation{Departement Fysica,
Universiteit Antwerpen, Groenenborgerlaan 171, B-2020 Antwerpen,
Belgium}

%\date{\today}

\begin{abstract}

The energy spectrum of a one-electron quantum dot doped with a
single magnetic ion is studied in the presence of an external
magnetic field. The allowed cyclotron resonance (CR) transitions are
obtained together with their oscillator strength (OS) as function of
the magnetic field, the position of the magnetic ion, and the
quantum dot confinement strength. With increasing magnetic field a
ferromagnetic - antiferromagnetic transition is found that results
in clear signatures in the CR absorption. It leads to
discontinuities in the transition energies and the oscillator
strengths and an increase of the number of allowed transitions.

\end{abstract}


\maketitle

\section{Introduction.}
Twenty years after the pioneering studies\cite{Furdyna} on
magnetically doped quantum dots, both experimental and theoretical
work on III-V and II-VI semiconductor systems and containing a low
density of magnetic ions \cite{Klein,Chen} (Diluted Magnetic
Semiconductors - DMS) have revealed a variety of remarkable new
physical
properties\cite{Chang,Besombes,Govorov,Archer,Schmidt,Worjnar,Fernandez,
Leger,Savic,Cheng,Hawrylak,NgaTTNguyen,Nguyen} as e.g. giant Zeeman
splitting, spin splitting of exciton levels, magnetic polarons, etc.
Recent investigations of the optical properties of II-VI
manganese-doped quantum dots\cite{Besombes,Fernandez,Leger,Savic}
have shown  that they are promising new systems for e.g. qubits. The
spin of the magnetic ion (Mn-ion) is used as a quantum bit and
recent experiments have shown that in single Mn-doped one-electron
quantum dots it is possible to control the spin of the electron and
of the magnetic ion\cite{Leger}. The characterization of single
Mn-doped few-electron II-VI quantum dots is very important.
Typically, one uses inter-band absorption experiments. Here, we will
investigate intra-band transitions and show that as function of the
magnetic field new features appear when a magnetic ion is present
which depend on the position of the Mn ion inside the quantum dot
(QD).

Studies on $N_e$ strongly interacting electrons that are confined in
a quantum dot and interacting with a single magnetic ion have been
mostly limited to the experimental\cite{Leger} and theoretical
study\cite{Fernandez,Hawrylak,NgaTTNguyen,Nguyen} of the ground
state and the thermodynamic properties of dots containing a small
number of electrons.

Cyclotron resonance (CR) has been an important experimental
technique to investigate the properties of electrons (and holes).
The advantage is that CR transitions involve only one type of
carriers (electrons or holes) facilitating the interpretation of the
results. CR is also often used to obtain information on the
effective mass of the carriers.

Cyclotron resonance has been studied in bulk
semiconductors\cite{Kohn} and later on also in semiconductor quantum
dots\cite{Merkt,Geerinckx,Peeters} where it was shown that in
parabolic confined quantum dots Kohn's theorem still holds and thus
the CR-transitions are independent of the number of electrons. The
latter is no longer true if the confinement potential is
non-parabolic\cite{Geerinckx} or when the carriers obey a
non-parabolic energy spectrum.

Very recently\cite{Savic}, the intra-band optical properties of a
three-dimensional (3D) single-electron CdTe/ZnTe quantum dot
containing one or two Mn impurities confined by a potential that is
parabolic in the xy-plane and has the quantum-well confinement along
the z-axis was investigated. New absorption lines (mixing with the
lines to higher excited levels for quantum-well system - as for
example see Ref.\cite{Geerinckx} for studies of CR in quantum well
system without a Mn-ion) were predicted as well as crossings and
anti-crossings of allowed CR transitions as function of an external
magnetic field.

In the present paper we calculate the single-particle states of a
quasi-two-dimensional parabolic quantum dot in the presence of a
single Mn-ion and an external perpendicular magnetic field. We
investigate the allowed cyclotron resonance transitions and their
corresponding oscillator strength. We have in mind experimental
realized systems [i.e. Cd(Mn)Te] where the lateral size is much
larger than the height of the quantum dot and consequently the
system behaves like a quasi-two-dimensional quantum dot. As in
Ref.\cite{Savic} we find new resonant transitions due to the
electron-magnetic ion (e-Mn) spin-spin exchange interaction but we
also investigate these transitions as function of the position of
the Mn-ion in the quantum dot. We find crossing and anti-crossing
features that depend on the position of the magnetic ion in the
quantum dot and on the strength of the confinement potential. We
will focus on the allowed transitions and concentrate on the new CR
lines that become possible due to the presence of the Mn-ion. Due to
the presence of the Mn-ion the electron spin will not always be
parallel to the external magnetic field. It will together with the
Mn-ion define two kinds of relative orientation of their spins
called ferromagnetic (FM) and antiferromagnetic (AFM). The FM-AFM
transition can be influenced by changing the exchange interaction
parameters which we will realize by moving the magnetic ion to
different positions. We show that this can change the position and
the number of crossing and anti-crossing points and consequently
influence the CR lines. We find that the CR transitions now contain
information not only on the electron state but also on the
magnetic-ion-electron interaction and the magnetic state of the
magnetic ion. Because of the rich physics involved we will limit our
discussion here to the fundamental single-electron quantum dot case.

This paper is organized as follows. Section II introduces the model
and the numerical method. In section III we present our numerical
results for the energy spectrum. Sect. IV is devoted to the
cyclotron resonance transitions where we calculate the oscillator
strength and the allowed transition energy spectrum. Our discussion
and conclusions are presented in Sect. V.

\section{Theoretical model}
A quantum dot containing a single electron with spin
$\overrightarrow{s}$ confined by a parabolic potential and
interacting with a single magnetic ion ($Mn^{2+}$) with spin
$\overrightarrow{M}$ and a magnetic field is described by the
following Hamiltonian:
\begin{align} \label{e:Hamiltonian}
\hat{H} =&
\left[\frac{1}{2m^*}{\left(-i\hbar\overrightarrow{\nabla}_{\overrightarrow{r}}+
    e\overrightarrow{A}(\overrightarrow{r})\right)^2}
    + \frac{1}{2}{m^*\omega_0^2\overrightarrow{r}^2}\right]
\nonumber\\
&+ \frac{1}{2}\hbar\omega_c\left(g_e m^*
 {s}_z + g_{Mn} m^* {M}_z\right)
 \nonumber\\
&-J_c\overrightarrow{s}\cdot
 \overrightarrow{M} \delta(\overrightarrow{r}-\overrightarrow{R}).
\end{align}

The first two terms are, respectively, the single-particle kinetic
energy and the confinement potential for the electron. The third and
the fourth terms are the electron and magnetic ion Zeeman energies,
respectively. The last term is the electron-$\emph{Mn}$ spin-spin
exchange interaction with strength $J_c$. The vector potential is
taken in the symmetric gauge: $\overrightarrow{A}=B/2(-y,x,0)$ where
the magnetic field $\overrightarrow{B}$ is perpendicular to the
plane of the interface. The confinement frequency $\omega_0$ defines
the confinement length: $l_0=\sqrt{{\hbar}/{m^*\omega_0}}$. $g_e$
and $g_{Mn}$ are the Land\'{e} g-factor of the host semiconductor
and the magnetic ion, respectively. We introduce a dimensionless
parameter $S_C=\left({a_B^{*}/l_0}\right)^2$ called the confinement
strength (square comes from the fact that $\omega_0\sim1/l_0^2)$. In
a many-electron system $S_C$ is: $S_C=1/\lambda_C^2$ with
$\lambda_C=l_0/a_B^{*}$ the Coulomb interaction
strength\cite{NgaTTNguyen}.
$a_B^{*}=4\pi\epsilon_0\epsilon\hbar^2/m^*e^2$ is the effective Bohr
radius. The cyclotron frequency is: $\omega_c=eB/m^*$. We use the
set of parameters\cite{Hawrylak} that is applicable to a
self-assembled Cd(Mn)Te quantum dot with a typical lateral size of
about tens of nanometers. The dielectric constant $\epsilon=10.6$,
effective mass $m^*=0.106 m_0$, $a_B^{*}=52.9$ ${\AA}$, $g_e=-1.67$,
$g_{Mn}=2.02$, $J_c=1.5\times 10^{3} meV {\AA}^{2}$, and $l_0$ about
tens of angstroms ($\hbar\omega_0$ corresponding to tens of $meV$).
For example, with $\hbar\omega_0=51.32$ $meV$ gives $l_0=26.45$
{\AA} and corresponds to $S_C=4$; $\hbar\omega_0=12.83$ $meV$ gives
$l_0=52.9$ {\AA} and corresponds to $S_C=1$.

We use the single-particle states in a parabolic confinement
potential namely the complete basis of Fock-Darwin (FD) orbitals
$\phi_{nl}\left(\overrightarrow{r}\right)$ and spin functions
$\chi_{\sigma}\left(\overrightarrow{s}\right)$:
\begin{equation}\label{e:basis}
\phi_{nl\sigma}\left(\overrightarrow{r},\overrightarrow{s}\right)=
\varphi_{nl}\left(\overrightarrow{r}\right)\chi_{\sigma}\left(\overrightarrow{s}\right),
\end{equation}
with the Fock-Darwin orbitals:
\begin{equation}\label{e:Fock-Darwin}
\varphi_{nl}\left(\overrightarrow{r}\right)=
\frac{1}{l_H}\sqrt{\frac{n!}{\pi\left(n+|l|\right)!}}\left(\frac{r}{l_H}\right)^{|l|}
e^{-il\theta}e^{-\frac{r^2}{2l_H^2}}L_n^{|l|}\left(\frac{r^2}{l_H^2}\right).
\end{equation}

The single-particle (spin $\sigma$) Fock-Darwin orbital energy in
the presence of a magnetic field at site $i=\{n,l\}$ is given by:
\begin{equation}
E_{i,\sigma}=\hbar\omega_H (2n+|l|+1)-\hbar\omega_cl/2,
\end{equation}
with  $\omega_H=\omega_0\sqrt{1+({{\omega_c}/{2\omega_0}})^2}$ that
can also be expressed in terms of a new length $l_H$:
$\omega_H=\hbar/m^*l^2_H$.

We now can rewrite the Hamiltonian in second-quantized form:
\begin{align}\label{e:secondquantized}
\hat H =& \sum_{i,\sigma}E_{i,\sigma}c_{i,\sigma}^+c_{i,\sigma}
+\frac{1}{2}\hbar\omega_c\left(g_e m^*
 {s}_z+g_{Mn} m^*{M}_z\right)
 \nonumber\\
 &- \sum_{ij}^{} \frac{1}{2}{J_{ij}\left(\overrightarrow{R}\right)}
 [(c_{i,\uparrow}^+
 c_{j,\uparrow}-c_{i,\downarrow}^+c_{j,\downarrow})M_z
 \nonumber\\
 &+ c_{i,\uparrow}^+c_{j,\downarrow}M^{-}
 + c_{i,\downarrow}^{+} c_{j,\uparrow} M^{+}],
\end{align}
where $i$ denotes a set of quantum numbers $\{n,l\}$. $L_z$, $s_z$,
and $M_z$ are the projections of the angular momentum of the
electron, its spin, and spin of the magnetic ion in the direction of
the magnetic field, respectively. $M_z, M^{+}$, and $M^{-}$ are the
z-projection, raising and lowering operators of the magnetic ion
spin, respectively. The e-Mn interaction appearing in this form
describes the conservation of the electron spin in the e-Mn
interaction of e.g. two configurations $i$ and $j$ in the first term
and the spin exchange of these configurations by flipping the
electron's spin at one site along with a compensation by flipping
the Mn-ion spin at the other site, and vice versa, in the last two
terms. This interaction parameter is:
\begin{equation}\label{e:JJex}
J_{ij}(\overrightarrow{R})=J_c
\varphi_i^*(\overrightarrow{R})\varphi_j(\overrightarrow{R})
\end{equation}
a product of the two Fock-Darwin orbitals of states $i$ and $j$
calculated at the position of the magnetic ion $R_{Mn}$.

We construct the many-particle wave function  using the
configuration interaction (CI) method:
\begin{equation}\label{e:Psi}
\Psi\left(\overrightarrow{x^*},
\overrightarrow{M}\right)=\sum^{N_{c}}_{k=1} c_k \Psi_{k}
\left(\overrightarrow{x^*}, \overrightarrow{M}\right),
\end{equation}
where $\overrightarrow{x^*}$ is the radial and spin coordinate of
the electron and $\Psi_k$ is the $k$-th state of the non-interacting
many-electron wave function (in this case for one-electron)
determined by:
\begin{equation}\label{e:Psi_k}
\Psi_k \Longrightarrow |k\rangle =
|c^+_{{k\uparrow}(\downarrow)}\rangle|M_z^{k}\rangle.
\end{equation}
$M_z^{k}$ runs through six states: $5/2$, $3/2$, ... $-5/2$. The
number of configurations, $N_{c}=12N_s$ with $N_s$ the number of
orbitals that are included and is taken sufficient large in order to
guarantee sufficient accuracy.

\section{Energy spectrum}
\begin{figure}[btp]
\begin{center}
\vspace*{-0.5cm}
\includegraphics*[width=9.2cm]{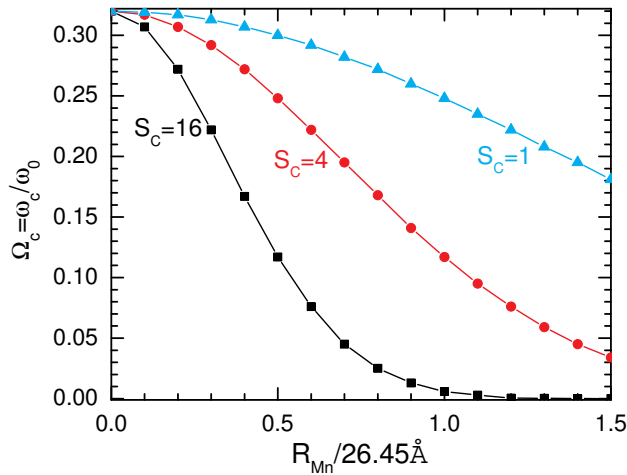}
\end{center}
\vspace{-0.5cm} \caption{(Color online) Phase diagram
$\Omega_C-R_{Mn}$ plotted for three different confinement strengths
$S_C$. The region below (above) the line corresponds each time to
the ferromagnetic (antiferromagnetic) phase, respectively. }
\label{PD_Ne=1}
\end{figure}

The presence of the magnetic ion leads to different arrangements of
the spins of the electron and the magnetic ion depending on the
strength of the magnetic field. This is due to the interplay between
on the one hand the contributions of the Zeeman energies of the
electron and the magnetic ion and on the other hand the exchange
interaction, both depend on the magnetic field. Two different
arrangements of the spins of the electron and the Mn-ion are
possible: ferromagnetic and antiferromagnetic. The ferromagnetic
arrangement is found for very small magnetic field when the exchange
energy dominates over the Zeeman energy terms. In the reverse case
the AFM arrangement of electron and magnetic ion spins is found. As
illustrated in Fig.~\ref{PD_Ne=1}, the ferromagnetic -
antiferromagnetic (FM-AFM) phase transition line depends also on the
position of the magnetic ion. Note that in nonzero magnetic field,
the magnetic ion is always antiparallel to the magnetic field due to
the positive value of its $g$-factor.

When the magnetic ion is located at the center of the dot the FM-AFM
transition occurs at the same magnetic field for any value of the
confinement strength as is made clear in Fig.~\ref{PD_Ne=1}. The
e-Mn exchange energy is maximal for a Mn-ion situated in the center
of the quantum dot. Note that in one-electron system, the electron
density is maximum at the center of the dot and the electron is
found in the $s$ shell where the $J_c$ matrix element is also found
to be maximum at the center of the dot. Moving the Mn-ion out of the
center of the quantum dot will decrease this e-Mn exchange
interaction. The electron probability at the Mn-ion is now a strong
function of the confinement strength. With decreasing e-Mn exchange
energy (i.e. by moving the Mn-ion out the center of the QD or
increasing the confinement) a smaller external magnetic field is
needed to induce the FM-AFM transition, as can be seen from
Fig.~\ref{PD_Ne=1}.
\begin{figure*}[btp]
\begin{center}
\vspace*{-0.5cm}
\includegraphics*[width=16.cm]{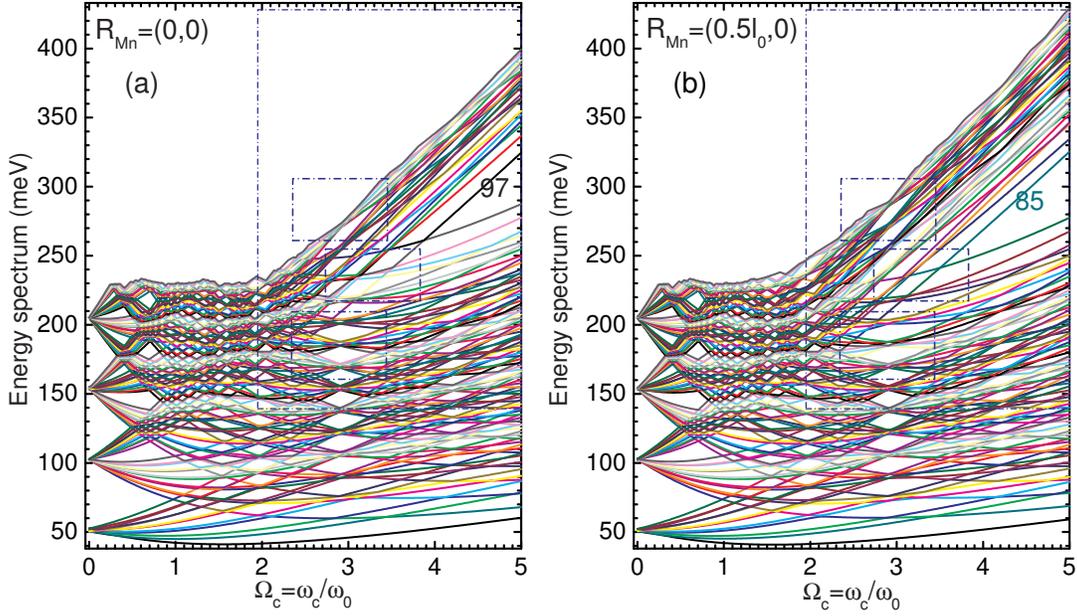}
\end{center}
\vspace{-0.5cm} \caption{(Color online) First $120$ levels of the
energy spectrum of a single-electron quantum dot with the magnetic
ion located at the center (a) and at $(0.5l_0,0)$ (b) of the QD for
confinement strength $S_C=4$. The colors that match the
corresponding energy levels are kept the same in the two plots. The
blue dash-dotted boxes highlight the regions where both figures
exhibit the largest difference. } \label{Spec_cen+0.5l0_Ne=1}
\end{figure*}

In the presence of the exchange energy term in Eq.
(\ref{e:Hamiltonian}) the ground state (GS) consists of
contributions from different configurations of ($s_z,M_z$). The
electron is still found in the $s$ shell even at very high magnetic
field. Unlike the system without a magnetic ion where the spin of
the electron always orients parallel to the magnetic field, the
system containing a magnetic ion has a wave function containing
contributions from states where the electron is antiparallel to the
magnetic field. This leads to an attraction to the magnetic ion and
reduces the total energy.

When a magnetic ion is placed inside the quantum dot the energy
spectrum is modified to the one given in
Fig.~\ref{Spec_cen+0.5l0_Ne=1} which shows the results in the case
the Mn-ion is situated at the center
[Fig.~\ref{Spec_cen+0.5l0_Ne=1}(a)] and at $(0.5l_0,0)$
[Fig.~\ref{Spec_cen+0.5l0_Ne=1}(b)] inside the quantum dot. Note
that now each $B=0$ energy level is split into many different levels
having different Zeeman splitting. There are also many more
crossings and anti-crossings and periodic opening of energy gaps in
the spectra. The position and the number of energy gaps, and
consequently the crossing and anti-crossing points, depend on the
position of the magnetic ion as can be seen in
Fig.~\ref{Spec_cen+0.5l0_Ne=1} in particular the regions inside the
blue boxes. We explicitly refer to energy levels $85$
[Fig.~\ref{Spec_cen+0.5l0_Ne=1}(a)] and $97$
[Fig.~\ref{Spec_cen+0.5l0_Ne=1}(b)] to highlight the differences in
the energy spectrum. When changing the position of the Mn-ion we
find that some of the crossing points turn into anti-crossing points
and at some of the anti-crossings the size of the energy gap
increases. That will be shown and discussed in the following
section.

The $B=0$ ground state level is split into $12$ levels because it is
composed of $s_z=\pm1/2$ and $M_z=\pm5/2, \pm3/2, \pm1/2$ spin
states. The next $B=0$ level is the $p$-electron state and because
the electron orbital momentum is $1$ this level splits into $24$
levels. Note that for small magnetic fields, the states whose wave
function have the $(0,1)$ or the $(0,-1)$ Fock-Darwin states as
their largest contribution stay lower in energy as compared to e.g.
the states whose wave function have $(0,2)$ or $(1,0)$ etc as the
largest contribution. However, as the magnetic field increases,
there are changes in relative position of the energy levels, i.e.
whose wave function has $(0,-1)$ as the largest contribution will
now be higher in energy than e.g. the state which has $(0,2)$ as the
major contributing Fock-Darwin state.
\begin{figure}[btp]
\begin{center}
\vspace*{-0.5cm}
\includegraphics*[width=9.2cm]{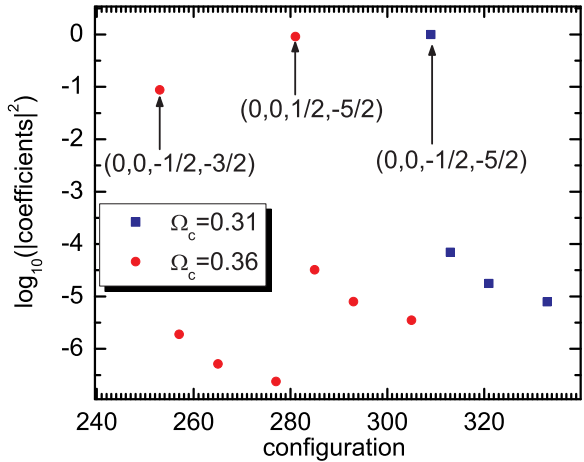}
\end{center}
\vspace{-0.5cm} \caption{(Color online) Contributions to the
ground-state wave function from the different Fock-Darwin wave
functions (x-axis) in case of the quantum dot of
Fig.~\ref{Spec_cen+0.5l0_Ne=1}(a) for two different magnetic fields:
$\Omega_c=0.31$ (FM) and $\Omega_c=0.36$ (AFM).} \label{GSWF_Ne=1}
\end{figure}

The presence of the magnetic ion leads to a mixing of the
Fock-Darwin orbitals for the different eigenstates in contrast to
the quantum dot without a magnetic ion where those Fock-Darwin
orbitals are the eigenstates of the dot. We show in
Fig.~\ref{GSWF_Ne=1} the contribution of the different Fock-Darwin
states to the wave function of the GS for two different values of
the magnetic field: $\Omega_c=0.31, 0.36$. $N_c$ configurations that
were defined in Sec. II are the x-axis. The configurations that
contribute more than $1\%$ are indicated by four quantum numbers
($n,l,s_z,M_z$). At $\Omega_c=0.31$ where the system is in the FM
phase, the GS wave function has the Fock-Darwin state
(0,0,-1/2,-5/2) as the largest contribution while at AFM magnetic
field $\Omega_c=0.36$ the two states $(0,0,-1/2,-3/2)$ and
$(0,0,1/2,-5/2)$ give the dominant contribution.
\begin{figure}[btp]
\begin{center}
\vspace*{-0.5cm}
\includegraphics*[width=9.2cm]{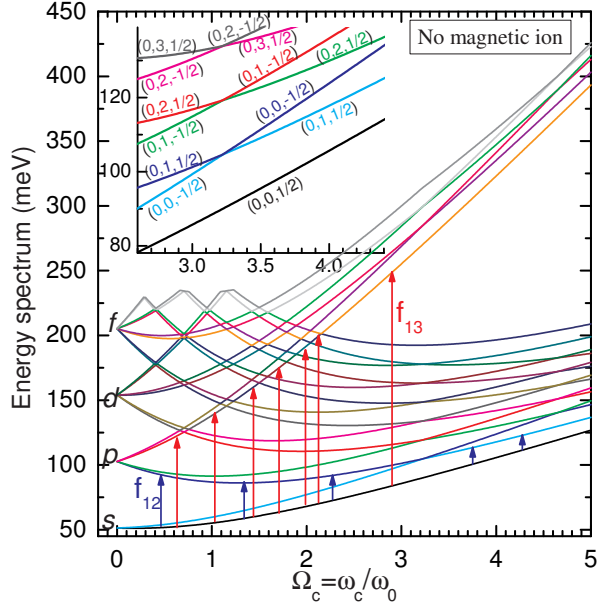}
\end{center}
\vspace{-0.5cm} \caption{(Color online) First $20$ levels (up to the
$f$-shell) of the energy spectrum of a one-electron quantum dot
without a magnetic ion. The inset is a zoom of the region
$2.6<\Omega_c<4.4$ that clarifies the CR-transitions of the electron
from the GS $(0,0,1/2)$ to either of $(0,1,1/2)$ and $(0,-1,1/2)$
with oscillator strengths $f_{12}$ (blue) and $f_{13}$ (red),
respectively. The energy levels are indicated by their quantum
numbers $(n,l,s_z)$. The allowed transitions are indicated by the
blue and red arrow.} \label{spectrum_UC}
\end{figure}

\section{Cyclotron transitions and oscillator strength}
In a cyclotron resonance experiment, an electron in quantum state
$i=(n,l)$ with energy $E_{i}$ is excited to a higher energy state
$E_{j}$ [$j=(n',l')$] with transition amplitude $A_{ij}$, the
associated oscillator strength for circular polarized light
is\cite{Geerinckx}:
\begin{equation}\label{e:OS}
 f_{ij}=\frac{2\Delta
 E_{ij}}{\hbar \omega_H}\cdot\frac{|A_{ij}|^2}{l_H^2},
\end{equation}
where
\begin{equation}\label{e:Aij}
 A_{ij}=<\Psi_i(\overrightarrow{x^*},\overrightarrow{M})|r
 e^{\pm i\theta}|\Psi_j(\overrightarrow{x^*},\overrightarrow{M})>.
\end{equation}
$\Delta E_{ij}=E_j-E_i$ is the transition energy. It is clear that
the state of the magnetic ion is not altered during such a
transition. The electron transition fulfills the selection rules:
$\Delta l=l'-l=\pm1$; $\Delta s_z=0$; and $\Delta M_z=0$.
\begin{figure}[btp]
\begin{center}
\vspace*{-0.5cm}
\includegraphics*[width=9.2cm]{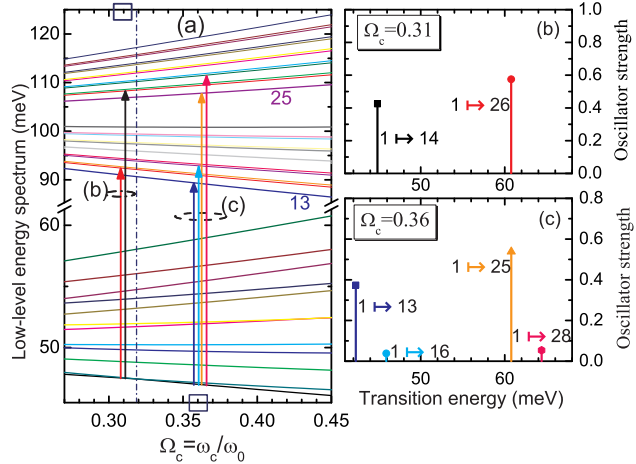}
\end{center}
\vspace{-0.5cm} \caption{(Color online) (a) Low energy spectrum
where some of the allowed transitions are indicated by arrows around
the magnetic fields $\Omega_c=0.31$ and $0.36$. The corresponding
oscillator strengths of these transitions are shown in (b) and (c).
The magnetic impurity is located at the center of the dot and the
confinement strength $S_C=4$. The blue dash-dotted line at
$\Omega_c=0.32$ indicates the ground-state FM-AFM transition.}
\label{Spec+OS_cen_Ne=1_B0.31}
\end{figure}

A general form for the wave function of state $i$ is:
\begin{equation}
\Psi_i(\overrightarrow{x^*},\overrightarrow{M})=\sum_{\alpha}^{N_c}
c_{\alpha}\Psi_{i\alpha}(\overrightarrow{x^*},\overrightarrow{M}),
\end{equation}
where $\Psi_{i\alpha}(\overrightarrow{x^*},\overrightarrow{M})$ is
the single-electron Fock-Darwin solution in the presence of the
magnetic ion defined via Eq. (\ref{e:Psi_k}). Now we calculate
$A_{ij}$ by integrating Eq. (\ref{e:Aij}) over $\overrightarrow{r}$.
The final expression for $A_{ij}$ can be written as:
\begin{equation}\label{e:Aif_final}
A_{ij}=\sum_{\alpha}^{N_c}\sum_{\beta}^{N_c}c_{\alpha}^{*}c_{\beta}
\delta_{{s_{z_{\alpha}}}{s_{z_{\beta}}}}
\delta_{M_{z_{\alpha}}M_{z_{\beta}}}A_{ij}^{\alpha \beta},
\end{equation}
where
\begin{eqnarray}
A_{ij}^{\alpha
\beta}=\delta_{n_{\beta},n_{\alpha}}\delta_{l_{\beta},l_{\alpha}\pm1}l_H
\sqrt{n_{\alpha}+|l_{\alpha}|+1}
\nonumber\\
-\delta_{n_{\beta},n_{\alpha}+1}\delta_{l_{\beta},l_{\alpha}\pm1}
(1-\delta_{l_{\alpha},0})l_H\sqrt{n_{\alpha}+1}
\end{eqnarray}
is the transition amplitude element identical to Eq. (18) of
Ref.\cite{Geerinckx}. The transition amplitude (\ref{e:Aif_final})
is a sum over all possible transition amplitudes of the respective
single states $\alpha$ and $\beta$.
\begin{figure}[btp]
\begin{center}
\vspace*{-0.5cm}
\includegraphics*[width=9.2cm]{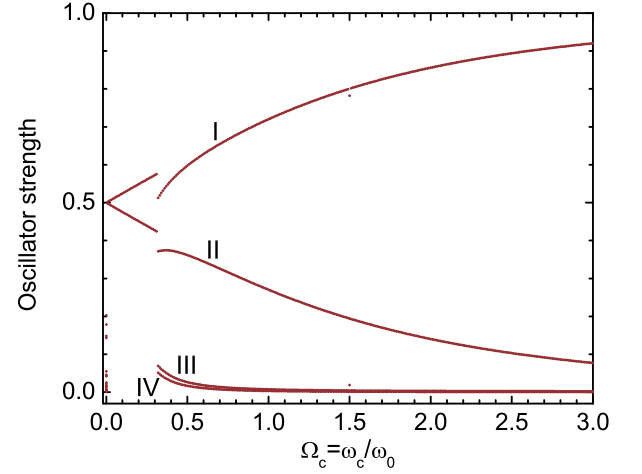}
\end{center}
\vspace{-0.5cm} \caption{(Color online) Oscillator strength of the
allowed electron transitions at various magnetic fields of the same
magnetic quantum dot as plotted in
Fig.~\ref{Spec+OS_cen_Ne=1_B0.31}. I, II, III, and IV indicate the
four branches of the OS curve as $\Omega_c\ge0.32$.}
\label{OS_cen_Ne=1_lamb1}
\end{figure}

To simplify the subsequent discussion we show in
Fig.~\ref{spectrum_UC} first the results for the case when no
magnetic ion is present. The selection rule allowed transitions
(i.e. $\Delta n=0$, $\Delta l=\pm1$) are indicated by the vertical
arrows which are transitions from the ground state $(0,0)$ to the
$p$-shell $(0,1)$ ($p^{+}$) and $(0,-1)$ ($p^{-}$) with the
respective oscillator strengths $f_{12}$ (blue) and $f_{13}$ (red).
As we can see from this figure, with increasing the magnetic field,
the states with positive $l$ have lower transition energy as
compared to the negative $l$. This is illustrated by the blue and
red arrows in Fig.~\ref{spectrum_UC}. From this figure, for the
magnetic field range $\Omega_c<3.22$, $f_{12}$ corresponds to the
transition of the electron from the ground state to the third state
(level $1\rightarrow3$) while the other transition $f_{13}$
corresponds to the transition of the electron from the ground-state
level to the fifth, and then seventh, and then ninth, and then
eleventh, and then thirteenth level. Note that in the limit of
$\Omega_c\rightarrow \infty$, $f_{12} \rightarrow 0$.
\begin{figure}[btp]
\begin{center}
\vspace*{-0.5cm}
\includegraphics*[width=9.2cm]{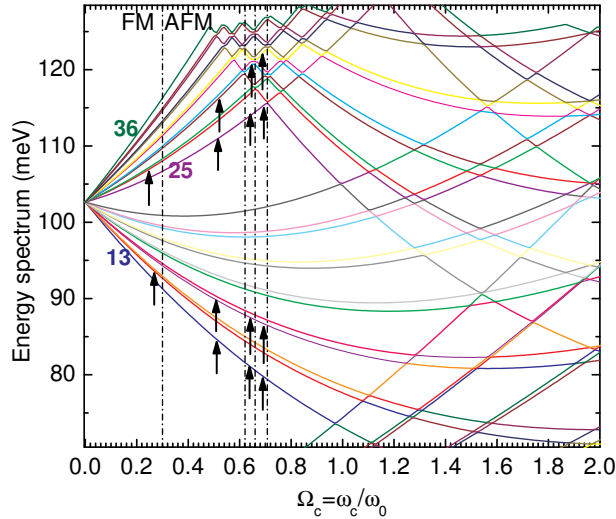}
\end{center}
\vspace{-0.5cm} \caption{(Color online) Low-level energy spectrum of
the same quantum dot as in Fig.~\ref{Spec+OS_cen_Ne=1_B0.31} is
plotted from the thirteenth to thirty-sixth levels to indicate the
final states of the electron transition. The black arrows indicate
the specific energy levels as the final state of the transition. The
final states shift following the arrows as the magnetic field
increases. The numbers: $13$, $25$, and $36$ are the levels' numbers
of the edge of subparts of the spectrum.}
\label{final_cen_Ne=1_lamb1}
\end{figure}

In the presence of the magnetic ion, due to the ferromagnetic (at
very small magnetic field) and antiferromagnetic (at larger magnetic
field) coupling of the electron with the magnetic ion, the energy
spectrum splits into many energy levels (see
Fig.~\ref{Spec_cen+0.5l0_Ne=1}) and no simple selection rules hold.
The CR spectrum now consists of many more peaks.
\begin{figure}[btp]
\begin{center}
\vspace*{-0.5cm}
\includegraphics*[width=9.2cm]{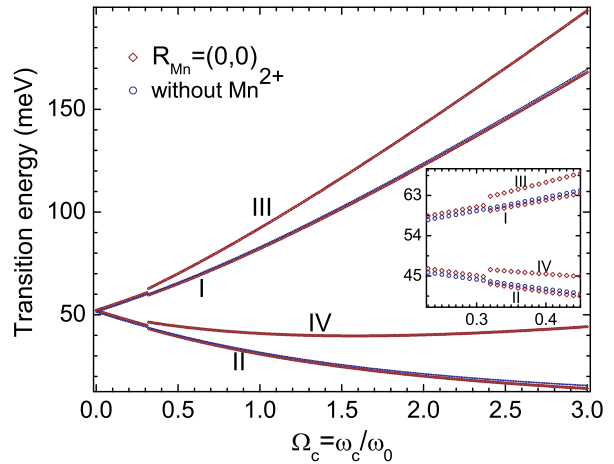}
\end{center}
\vspace{-0.5cm} \caption{(Color online) Transition energy for the
same quantum dot as plotted in Fig.~\ref{OS_cen_Ne=1_lamb1} with
four branches I, II, III, and IV corresponding to the four branches
in the OS curve. Blue curves are the transition energy in the
absence of a magnetic ion. The inset is a zoom around the FM-AFM
transition magnetic field.} \label{ERGTRAN_cen_Ne=1_lamb1}
\end{figure}

We first discuss in detail the electron transition for the case the
magnetic ion is located at the center of the quantum dot
[$R_{Mn}=(0,0)$] for magnetic fields $\Omega_c=0.31$ (ferromagnetic
phase) and $\Omega_c=0.36$ (antiferromagnetic phase - as can be seen
from Fig.~\ref{PD_Ne=1}). We found two allowed transitions as seen
in Fig.~\ref{Spec+OS_cen_Ne=1_B0.31} for $\Omega_c=0.31$ with their
respective resonant frequencies. Remember that at this magnetic
field (within the region $\Omega_c<0.32$), the system is in the
ferromagnetic phase where the electron and the magnetic ion are both
anti-parallel to the magnetic field. The transitions from the ground
state (level number $1$) where the electron is mostly in the $s$
orbital, to the state whose wave function has the orbital $(0,1)$ as
the largest Fock-Darwin contribution corresponding to the black
line; and to the state whose wave function has the orbital $(0,-1)$
as the largest contribution corresponding to the red line,
respectively. The resonances are found for transitions to the
fourteenth and the twenty-sixth level. The corresponding result for
the antiferromagnetic phase are shown in
Fig.~\ref{Spec+OS_cen_Ne=1_B0.31}(c) for $\Omega_c=0.36$. Notice
that now we have four possible transitions with oscillator strength
larger than $1\%$ instead of two in previous case. This is in
agreement with the effect found in Ref.\cite{Savic} These four
transitions correspond to the excitations of the electron from the
ground state to the energy levels: thirteenth (blue square),
sixteenth (cyan rhombus), twenty-fifth (orange triangle) and
twenty-eighth (pink circle). The first two: blue square and cyan
rhombus belong to the transition of the electron from the $s$ ground
state to the $p^{+}$ state (with positive angular momentum); while
the other two belong to the other transition of the electron to the
$p^{-}$ state (with negative angular momentum). The origin of this
difference as compared to the case for $\Omega_c=0.31$ can also be
explained by looking at the GS wave function contributions of the
Fock-Darwin states as shown in Fig.~\ref{GSWF_Ne=1}.
\begin{figure}[btp]
\begin{center}
\vspace*{-0.5cm}
\includegraphics*[width=9.2cm]{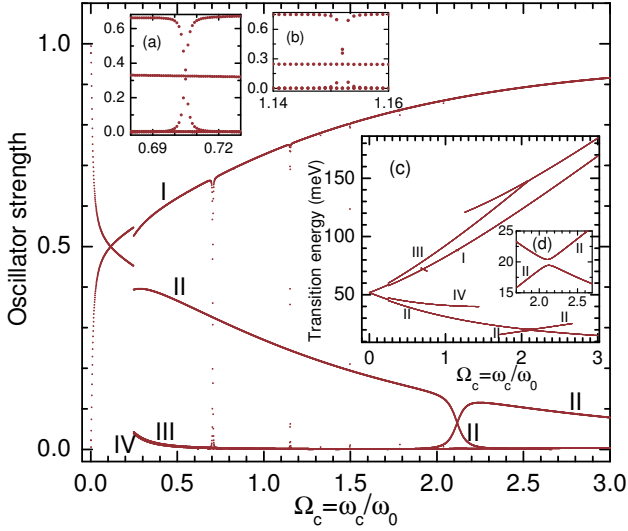}
\end{center}
\vspace{-0.5cm} \caption{(Color online) Oscillator strength with two
zooms around $\Omega_c=0.71$ (a) and $\Omega_c=1.15$ (b) and
transition energy (c) with a zoom around $\Omega_c=2.12$ (d) for the
same quantum dot as plotted in Fig.~\ref{OS_cen_Ne=1_lamb1} but the
magnetic ion located at $(0.5l_0,0)$. I, II, III, and IV indicate
the four branches when $\Omega_c\ge0.25$ in the OS and the
corresponding transition energies. }
\label{OS+Etran_0.5l0_Ne=1_lamb1}
\end{figure}

From Fig.~\ref{OS_cen_Ne=1_lamb1} we find that the main transitions
come from transitions to the two $p$-orbitals. The oscillator
strength exhibits a discontinuity at $\Omega_c=0.32$ where the
FM-AFM transition takes place. The wave function of the ground state
changes from favoring the configurations with electron spin down to
the configurations with electron spin up as can be seen from
Fig.~\ref{GSWF_Ne=1}. As the magnetic field increases from
$\Omega_c=0.31$ where the system is in the FM phase to
$\Omega_c=0.36$ where the system is in the AFM phase, the largest
Fock-Darwin contribution to the GS wave function goes from
$(n,l,s_z,M_z)=(0,0,-1/2,-5/2)$ to $(0,0,-1/2,-3/2)$ and
$(0,0,1/2,-5/2)$. Consequently, the final state of the electron
transition shifts. We will clarify this point in
Fig.~\ref{final_cen_Ne=1_lamb1} by using the black arrows to direct
the attention of the reader to the final states for the transitions
of the electron. For example, for $\Omega_c<0.32$, the nonzero
resonances are the lines from the ground state to the fourteenth and
the twenty-sixth levels. When the system transits to the AFM phase,
the number of resonance lines increases from two to four. In the
region $0.32\leq \Omega_c < 0.62$, the transitions are to the
thirteenth and sixteenth level; the twenty-fifth and twenty-eighth;
and the main contributions are the lines to the thirteenth and the
twenty-fifth, etc. This is explained as follows. For the FM magnetic
field range $\Omega_c<0.32$, the electron and the Mn-ion both are
antiparallel to the magnetic field and $s_z+M_z=-3$ with the
configuration of $(n=0,l=0,s_z=-1/2, M_z=-5/2)$ having the
probability almost unity. This is due to the fact that the
commutator of the z-component of the total spin commutes with the
Hamiltonian, $[s_z+M_z,H]=0$. Beyond this magnetic field region,
$\Omega_c\geq 0.32$, the electron and the Mn-ion are antiparallel
and since the Zeeman (spin) part of the Mn-ion is always larger than
the electron's and the exchange energy becomes the smallest one
among these three competing energies there are two configurations
$(0,0,1/2,-5/2)$ and $(0,0,-1/2,-3/2)$ with $s_z+M_z=-2$ of the GS
wave function as the main contributions. It is obvious that the
exchange energy now has the largest contribution coming from the
second term [corresponding to configuration $(0,0,1/2,-5/2)$] and
the second largest from the last term [corresponding to
configuration $(0,0,-1/2,-3/2)$] in the last sum in Eq.
(\ref{e:secondquantized}). As the system is in the AFM, the final
states of the major parts in the OS that are the branches I and II
in Fig.~\ref{OS_cen_Ne=1_lamb1} and the corresponding transition
energies in Fig.~\ref{ERGTRAN_cen_Ne=1_lamb1} have wave functions
that are the FD states $(0,-1,1/2,-5/2)$ and $(0,1,1/2,-5/2)$,
respectively, as the major contributions. The other two transitions
(III and IV) come from the spin-spin exchange interactions
corresponding to the final states that have wave functions
containing FD states, respectively, $(0,-1,-1/2,-3/2)$ and
$(0,1,-1/2,-3/2)$ as their major contributions. In general, the
largest one among all the small contributions can reach to about $5
\div 10\%$ of the total OS. With increasing magnetic field, these
smaller contributions to the OS decrease to zero.
\begin{figure}[btp]
\begin{center}
\vspace*{-0.5cm}
\includegraphics*[width=8.2cm]{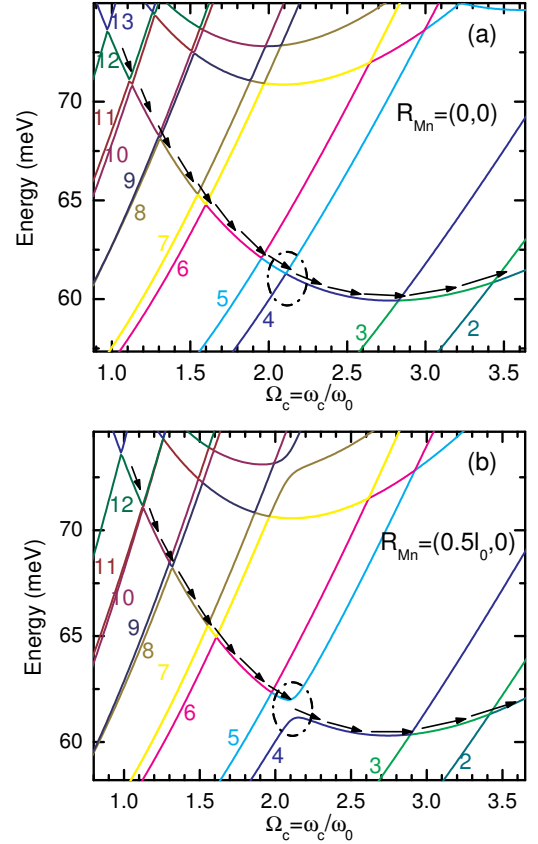}
\end{center}
\vspace{-0.5cm} \caption{(Color online) Zoom of the low-level
spectrum of Fig.~\ref{Spec_cen+0.5l0_Ne=1} for the cases the
magnetic ion is located at the center and at $(0.5l_0,0)$ in the
quantum dot. The black arrows indicate how the final state moves
with magnetic field for the lowest energy CR transition.}
\label{zoompoint2.1}
\end{figure}
\begin{figure}[btp]
\begin{center}
\vspace*{-0.5cm}
\includegraphics*[width=7.2cm]{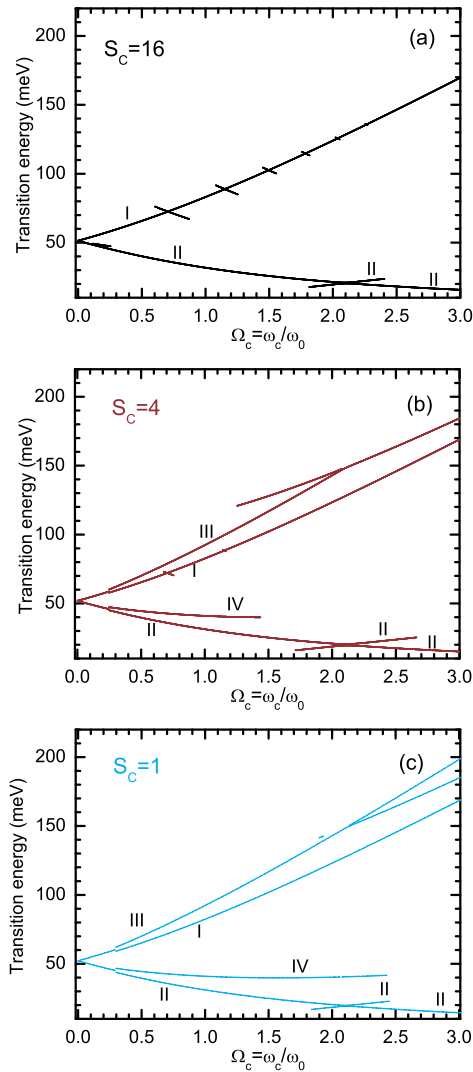}
\end{center}
\vspace{-0.5cm} \caption{(Color online) Transition energy for the
quantum dot with the magnetic ion located at $(0.5l_0,0)$ and for
three different values of $S_C$. I, II, III, and IV indicate the
four [two in (a)] main branches when the system is in the AFM phase.
The transition energy is scaled by the ratio of $S_C$ to $S_C=4$.}
\label{Etran_lamb0.5+1+2_Ne=1}
\end{figure}

The oscillator strength for the case the magnetic ion is located at
the center of the dot samples the central region of the quantum dot.
In this situation the magnetic ion does not interact with the
$p$-orbitals. The four branches in the OS
(Fig.~\ref{OS_cen_Ne=1_lamb1}) and transition energy
(Fig.~\ref{ERGTRAN_cen_Ne=1_lamb1}) in the AFM phase are smooth with
respect to magnetic field. With moving the Mn-ion to other positions
away from the center of the dot gives rise to the interactions of
the Mn-ion with the $p$-orbitals and this is expected to change the
OS by the changes in the two main branches (I and II in
Figs.~\ref{OS_cen_Ne=1_lamb1} and ~\ref{ERGTRAN_cen_Ne=1_lamb1}).
The influence of the position of the magnetic ion on the
CR-transitions is shown in Fig.~\ref{OS+Etran_0.5l0_Ne=1_lamb1} for
the case the Mn-ion is located e.g. at $(0.5l_0,0)$. First of all,
notice a discontinuity of the oscillator strength at $\Omega_c=0.25$
which corresponds to the FM-AFM transition. Second, we find an
unusual behavior at $\Omega_c=2.12$ where the lower curve exhibits a
crossing. Which is a consequence of the existence of an energy gap
between the fourth and the fifth levels at $\Omega_c=2.12$ as is
clearly shown in Fig.~\ref{zoompoint2.1}(b) for the case the Mn-ion
is located at $(0.5l_0,0)$. These fourth and fifth levels have the
Fock-Darwin states (0,1,1/2,-5/2) and (0,0,-1/2,-3/2) as their major
contributions. When the Mn-ion is located at $(0.5l_0,0)$ the
exchange matrix elements between the $p$ orbitals are nonzero which
leads to the anti-crossing. This is the reason that the
corresponding transition energy [see
Figs.~\ref{OS+Etran_0.5l0_Ne=1_lamb1}(c), (d)] exhibits an
anti-crossing behavior at $\Omega_c=2.12$. The energy gap is closed
for the case the Mn-ion is located at the center of the QD [see
Fig.~\ref{zoompoint2.1}(a)]. This similar behavior takes place at
other magnetic fields such as: $\Omega_c=0.71$ [see
Fig.~\ref{OS+Etran_0.5l0_Ne=1_lamb1}(a)] and $\Omega_c=1.15$ [see
Fig.~\ref{OS+Etran_0.5l0_Ne=1_lamb1}(b)] for the other kind of
electron transitions: the transition to the final state with
negative azimuthal quantum number. This is due to the fact that
around the above magnetic fields there exist two final states
contributing to the electron transition with their wave functions
composed of the two major FD states $(0,-1,1/2,-5/2)$ and
$(0,0,1/2,-5/2)$. This results in a smaller region of the magnetic
field as compared to the previous case around $\Omega_c=2.12$.
\begin{figure}[btp]
\begin{center}
\vspace*{-0.5cm}
\includegraphics*[width=7.0cm]{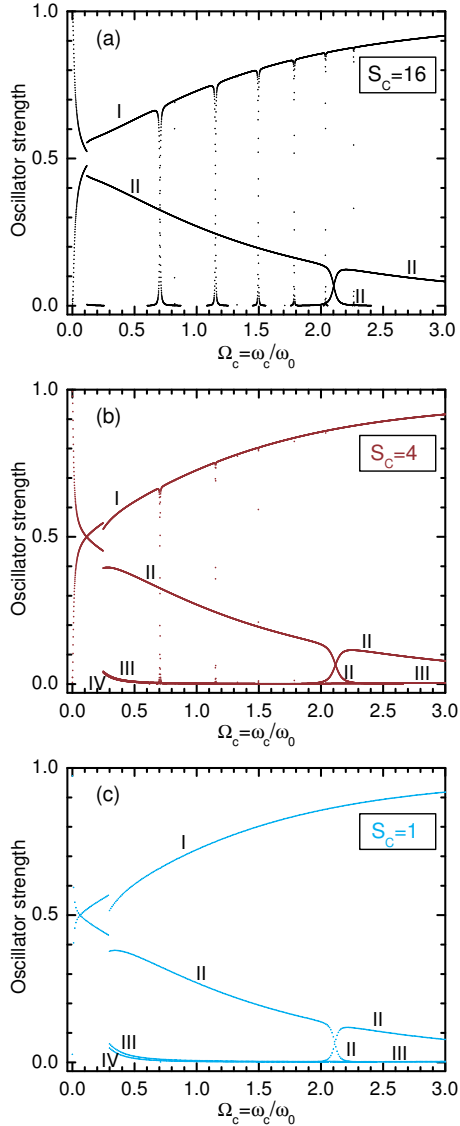}
\end{center}
\vspace{-0.5cm} \caption{(Color online) Oscillator strength with I,
II, III, and IV each time indicating the four main transitions [two
in (a)] corresponding to the four (three) branches in
Fig.~\ref{Etran_lamb0.5+1+2_Ne=1}.} \label{OS_lamb0.5+1+2_Ne=1}
\end{figure}

Next we discuss the confinement-strength dependence of the electron
transition first for the case that the Mn-ion is located at the
center of the quantum dot. We consider two different values of $S_C$
i.e. $S_C=4$ and $S_C=1$. The OS curve does not change with changing
the confinement strength while the transition energy is proportional
to $S_C$. This scaling is no longer exact if one displaces the
Mn-ion away from the center of the quantum dot, i.e. for
$R_{Mn}=(13.2{\AA},0)=(0.5l_0,0)$, as can be seen from
Fig.~\ref{Etran_lamb0.5+1+2_Ne=1}. The transition energies for three
different $S_C=16$, $4$, and $1$ that are all scaled to the case of
$S_C=4$ now exhibit different behaviors as seen in
Figs.~\ref{Etran_lamb0.5+1+2_Ne=1}(a), (b), and (c). First, the
FM-AFM transition that leads to a discontinuity is found at
different magnetic fields, i.e. $\Omega_c= 0.12$, $0.25$, and $0.3$
for $S_C=16$, $4$, and $1$, respectively. Besides, the anti-crossing
point at $\Omega_c=2.12$ as seen already from
Fig.~\ref{OS+Etran_0.5l0_Ne=1_lamb1} is different for different
$S_C$. It is now at $2.1$, $2.12$, and $2.11$ for $S_C=16, 4$, and
$1$, respectively. The OS also exhibits different features at the
above magnetic fields as can be seen from
Figs.~\ref{OS_lamb0.5+1+2_Ne=1}(a), (b), and (c). As the electron is
strongly confined, i.e. the case of $S_C=16$, the wave function
becomes strongly localized and because the Mn-ion is now located at
$(0.5l_0,0)$, that is far enough from the center of the dot, the
exchange interactions between the $p$-electron and the Mn-ion
reduce. Consequently, the effect of the e-Mn-ion interaction on the
CR-lines results in several anti-crossings in the transition energy
(crossings in the OS) as can be seen from
Figs.~\ref{Etran_lamb0.5+1+2_Ne=1}(a) and
~\ref{OS_lamb0.5+1+2_Ne=1}(a). The branches I and II in all six
figures from Fig.~\ref{Etran_lamb0.5+1+2_Ne=1}(a) to
Fig.~\ref{OS_lamb0.5+1+2_Ne=1}(c) refer to the transitions of the
electron from the GS to the final states which have the FD states
$(0,-1,1/2,-5/2)$ and $(0,1,1/2,-5/2)$ as the major contributions,
respectively.
\begin{figure}[btp]
\begin{center}
\vspace*{-0.5cm}
\includegraphics*[width=8.2cm]{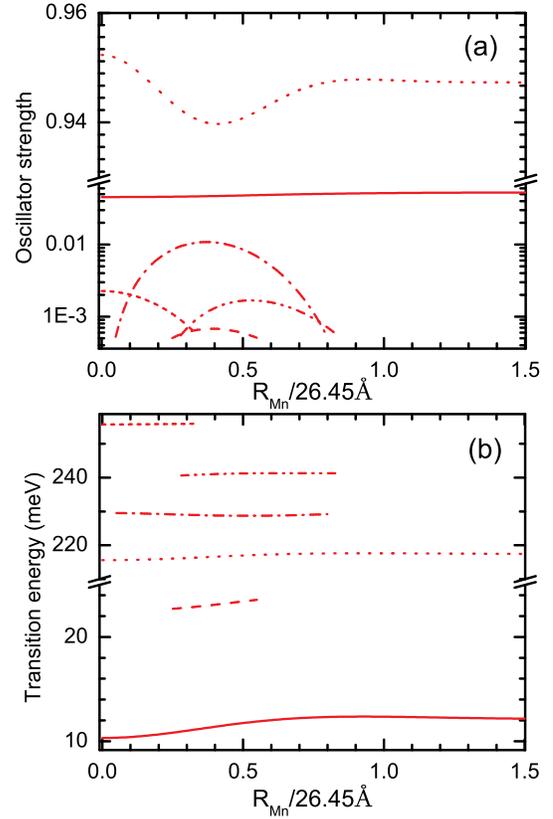}
\end{center}
\vspace{-0.5cm} \caption{(Color online) (a) The logarithm of the OS
and (b) the transition energy vs. the position of the magnetic ion
at magnetic field $\Omega_c=4$ for confinement strength $S_C=4$.
Solid and dotted lines are, respectively, the contributions from the
second and the ninety-seventh level as the main contributions to the
electron transitions to the two $p$ orbitals. } \label{OS_POS_Ne=1}
\end{figure}

We now explore the magnetic ion position dependence of the OS within
a large range of $R_{Mn}$. We realize that at high magnetic field,
the single-energy states ``convert" to the Landau levels with the
Landau indices $N_{L}=n+\frac{|l|-l+1}{2}$. We focus now on the high
magnetic field transitions for different positions of the Mn-ion and
different confinement strengths. As an example see
Fig.~\ref{OS_POS_Ne=1}, where at high magnetic field, e.g.
$\Omega_c=4$, the electron transitions are mainly between the GS to
the second (solid) and the ninety-seventh (dotted) level, the upper
part (dotted) of the OS curve oscillates and exhibits a maximum in
case the Mn-ion is at the center of the dot. In the latter case we
have the largest contribution of the exchange energy to the total
energy as compared to the other positions. Note also that if one
keeps increasing the magnetic field, the small-contribution branches
(dashed, dash-dotted, dash-dot-dotted, and short-dashed
corresponding to the $5^{th}$, $98^{th}$, $101^{st}$, and $105^{th}$
level, respectively) in the OS curve, which are generally only about
$2-5\%$ of their dominant branches (solid and dotted), become
smaller. In the extremely high magnetic field limit, the first
electron transition, namely to the energy level with the Fock-Darwin
state $(0,1)$ dominant, becomes zero making the transition to the
level with the Fock-Darwin state $(0,-1)$ dominant, increases to
unity. Therefore, the small-contribution branches in the OS curve
plotted in Fig.~\ref{OS_POS_Ne=1}(a) can be neglected in
experimental measurements.
\begin{figure}[btp]
\begin{center}
\vspace*{-0.5cm}
\includegraphics*[width=8.2cm]{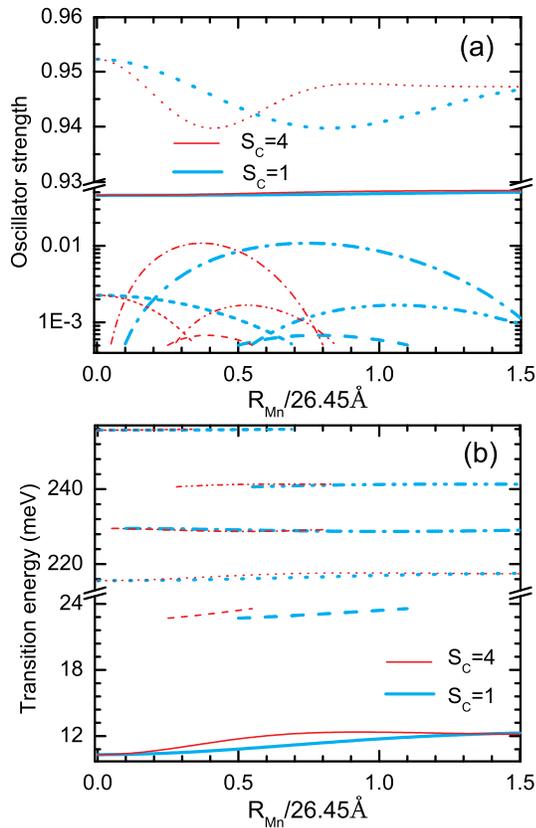}
\end{center}
\vspace{-0.5cm} \caption{(Color online) (a) The logarithm of the OS
and (b) the transition energies of the same QD as plotted in
Fig.~\ref{OS_POS_Ne=1} but for two different confinement strengths
$S_C=4$ (six red thinner curves) and $S_C=1$ (six cyan thicker
curves). The final levels with the same relative position in the OS
and two energy spectra (for two different $S_C$) have the same style
of line.} \label{OS_POS_Ne=1_severalSc}
\end{figure}

To complete this study we will examine the confinement-strength
dependence of the electron transition for various positions of the
Mn-ion. It is understandable that for a stronger confinement, the
electron becomes more localized resulting in a shifting of the OS
curve to a smaller $R_{Mn}$. This is shown in the OS behavior in
Fig.~\ref{OS_POS_Ne=1_severalSc}(a) where the thinner OS curve (with
six components) corresponds to the larger confinement strength
$S_C=4$. For less confined system (smaller value of $S_C$ i.e.
$S_C=1$), the small-contribution branches are more expanded. With
increasing $R_{Mn}$, the problem converts to the problem without a
magnetic ion. The small branches gradually disappear resulting in
the saturation of the oscillator strength and transition energy for
very large values of $R_{Mn}$.

\section{Discussions.}
We studied the CR transition properties of a $Mn^{2+}$-doped quantum
dot containing a single electron confined by a parabolic potential
in the presence of magnetic field. The transition energies and the
oscillator strength were obtained for different positions of the
magnetic ion with different confinement strengths.

As compared to the usual quantum dot (see e.g. Ref.\cite{Geerinckx})
without a magnetic ion, new transition frequencies are found to be
allowed. These new transitions are due to the presence of the spin
exchange interaction of the electron with the Mn ion which mixes the
Fock-Darwin orbitals. Furthermore, the energy spectrum exhibits many
anti-crossings of energy levels that are not present in the usual
single-electron parabolic quantum dot system.

At small magnetic fields the electron spin and the magnetic ion spin
are parallel which is similar to a FM state. With increasing
magnetic field a transition takes place where the electron spin
becomes parallel to the magnetic field direction leading to an AFM
state. This FM-AFM transition exhibits clear signatures in the CR
absorption spectrum: 1) the CR transition energies are
discontinuous, 2) the oscillator strength of the allowed transitions
are discontinuous, and 3) the number of allowed transitions, i.e.
the number of peaks in the CR absorption spectrum, is different in
the FM and in the AFM states. The magnetic field value at which the
FM-AFM transition occurs depends on the position of the magnetic ion
and on the confinement strength $S_C$ which also influences the CR
absorption spectrum. Besides, the anti-crossings in the energy
spectrum, which are due to the spin-spin exchange interactions,
result in anti-crossings in the transition energies and at the same
time crossings in the oscillator strengths. This is due to the fact
that around these anti-crossings (in the energy spectrum), there are
competitions between two energy levels as the major contribution to
the electron transition. Both of these energy levels contribute
about $40:60\%$ or $50:50\%$ to the transition e.g. from the GS to
the $p$ state with positive or negative azimuthal quantum number.
The number and the positions of these anti-crossings (crossings)
change with changing the Mn-ion position and $S_C$.

At high magnetic field where the energy levels ``convert" to Landau
levels, we obtained the magnetic-ion-position dependence of the
CR-lines for different confinement strengths and investigated it for
increasing confinement strength. The electron transition to higher
energy levels exhibits an oscillatory behavior as function of the
position of the magnetic ion. However, there is one particular ion
position, i.e. the center of the quantum dot, where the OS is
independent of the confinement strength.

\section{Acknowledgments}
We are thankful to Prof. A. Govorov for fruitful discussions. This
work was supported by FWO-Vl (Flemish Science Foundation), the EU
Network of Excellence: SANDiE, and the Belgian Science Policy (IAP).

%\begin{references}

\end{document}